\documentclass[conference]{IEEEtran}
\ifCLASSINFOpdf
\else
\usepackage[dvips]{graphicx}
\usepackage{epsfig}
\fi
\usepackage{algorithmic}
\usepackage{algorithm}

\usepackage{array}
\usepackage{url}

\begin{document}
%
\title{A Methodology for the Selection of Requirement Elicitation Techniques}



%
\author{

\IEEEauthorblockA{Saurabh Tiwari}
\IEEEauthorblockA{DA-IICT, Gandhinagar, India}
\IEEEauthorblockA{Email: saurabh$\textunderscore$t@daiict.ac.in}
\and
\IEEEauthorblockA{Santosh Singh Rathore}
\IEEEauthorblockA{Thapar University, Patiala, India}
\IEEEauthorblockA{Email: santosh.singh@thapar.edu}

}

\maketitle

\begin{abstract}
In this paper, we present an approach to select a subset of requirement elicitation technique for an optimum result in the requirement elicitation process. Our approach consists of three steps. First, we identify various attribute in three important dimensions namely project, people and the process of software development that can influence the outcome of an elicitation process. Second, we construct three p matrix (3PM) separately for each dimension, that shows a relation between the elicitation techniques and three dimensions of a software. Third, we provide a mapping criteria and use them in the selection of a subset of elicitation techniques. We demonstrate the applicability of the proposed approach using case studies to evaluate and provide the contextual knowledge of selecting requirement elicitation technique.
\end{abstract}

\begin{IEEEkeywords}
Requirement elicitation, elicitation techniques, elicitation technique selection, evaluation, framework.
\end{IEEEkeywords}

%
\IEEEpeerreviewmaketitle

\section{Introduction}
Software requirement elicitation is a fundamental and critical part of the software development life cycle. It is generally accepted that the quality of software depends on the requirements upon which software has been developed. The success or failure of a software development effort is greatly influenced by the quality of the requirements. Therefore, we require a variety of elicitation techniques beforehand to determine the user or customer needs. Though it is difficult to gather complete requirements from the users but choosing the best elicitation technique available in context with the software characteristics might ensure the completeness of requirements. In their study, BELL et al. [1] observed that, "The requirement for a system do not arise naturally; instead, they need to be engineered and have continuing review and revision".

It has been found that the cost of rectifying errors in software increases exponentially as one approaches the later phases of software development. Hence, the identification of missing and incorrect requirements in the requirement elicitation process itself would help to reduce the cost of rectifying errors in the software. If one identifies all incorrect and missing requirements then, theoretically the cost of rectifying errors later would approach zero. However, simultaneously the time taken to identify all such missing and incorrect requirements is infinite and hence, undertaking such a kind of task is impossible [73]. Thus, all we can do is, optimizing the process of requirement elicitation thereby doing it better and faster.

The selection of suitable requirement elicitation techniques for a specific domain of a software project is a challenging issue. In the literature, we identified that there exist numerous techniques of requirement elicitation to gather requirements but, due to the  lack of proper contextual knowledge of their use, it is difficult to choose the right technique [74]. There are a number of variables influences this selection process. Generally, the techniques are chosen by the requirement analysts for the specific project through their interest or experience. In the upcoming arena, there are numerous numbers of methods or techniques have been proposed by various authors to acquire information for the purposes of eliciting [19][21]. The motivation behind the study is to create a simple mapping platform between requirements elicitation techniques and threefold matrix consisting project, people and process attribute. The basic idea of mapping approach is taken from the astrology (\emph{Kundli}) software. The process of eliciting requirement, not only helps the organization to gather the requirements, but also in the analysis of requirement.

In this paper, we propose a framework to select effective elicitation techniques i.e., Choosing a small subset of requirements elicitation techniques from the set of techniques. Moreover, we focus to identify the project, people and process attributes for any problem domain and find out the relationship between these available three p matrix (3PM) with the available techniques to provide a guidance on selecting requirement elicitation technique. So, we could improve the average analyst's abilities to select a set of elicitation techniques, and we will most likely to improve the successfulness of software systems. Using all these matrix parameters, we applied a mapping mechanism on the framework for the evaluation and selection of requirement elicitation techniques. We have demonstrated the applicability of these matrices on the case studies and their comparison with the other elicitation based approaches. The results suggested that the matrices are capable of
selecting a set of effective elicitation techniques for a software in order to effectively elicit, model, document, verify and validate
requirements.

The rest of the paper is organized as follows. Section II, provides the background information about the requirement elicitation techniques.  Section III, presents the data collection strategy for the matrices. Section IV, describes an overview of the proposed framework. Section V, explains our approach and framework via case studies. Section VI, presents the comparison of our approach with other RE approaches. Subsequently, discuss the applicability of the proposed approach and other related issues in Section VII. We present related work in Section VIII, and finally summarize our conclusions in Section IX.

\section{Background Overview}\label{section3}
\subsection{Problem Description}
Requirement elicitation is a process of determining the problems and needs of the customer, so that software developers can construct a system that actually resolve customer problems and address their needs [33][34]. Understanding requirements is a difficult task because it involves natural language to interact with the end-users, and end users may provide incomplete and ambiguous requirements. Requirements are volatile in nature, and they change over the period of time. There are some social issues also, that affect requirement elicitation process. The process of selecting an elicitation technique is affected by a number of other parameters also. Often requirement analysts choose a technique based on the some assumptions [11]. That are:

\begin{itemize}
  \item This is the only technique which are they are known.
  \item Because this technique works effetely last time, so it will also work at this time.
  \item The analyst intuitively understands that the technique is effective in the current circumstance.
  \item The analyst is following some explicit methodology, and that methodology prescribes a particular technique at the current time.
\end{itemize}

Number of authors study about requirements elicitation problem [3][10][11][12][40], and they confirm that the problem is at much larger scale. Surveys revealed that one third of the projects started were never completed, and one half of them succeeded only partially. The reason behind  such failure is poor requirement elicitation - more precisely, the lack of user involvement, requirement incompleteness, ambiguity in requirements, unrealistic expectations and unclear objectives [17].

\subsection{Literature review: Techniques for Requirement Elicitation}
Many articles and books describe a way to perform requirement elicitation task. Most of the time practitioners are looking for a simple recipe that will solve all their elicitation problems. However, because of the nature of this problem, one elicitation technique cannot work in all situations. Therefore, the number of authors describes various requirement elicitation techniques [19][20][21][22][23][24][25].

Requirement elicitation techniques can be divided into four categories according to their nature of communication - traditional, contextual, collaborative and cognitive. The categorization of these requirement elicitation techniques refer from the Lecture of Requirement Elicitation at University of Toronto, department of computer science.

\subsubsection{Traditional Techniques}
\textbf{Interview} is a method of identifying facts and opinions of users and other stakeholders of the system under development by face to face conversation. There are two different kinds of interviews: The closed interview, where the requirements elector has a pre-defined set of questions and is looking for their answers. The open interview, where the requirements engineer and stakeholders discuss in an open-ended way to find out their expectation from a system.

\textbf{Questionnaire} is a technique of eliciting requirement from a large number of people in lesser cost and time. A well designed questionnaire can be useful to elicit the actual requirements from the stakeholders.

\textbf{Data gather from existing system} is used when we gather data for a system to replace an existing one. It is a useful technique to collect the depth knowledge of the system.

\textbf{Survey} is a technique of eliciting requirement from a large number of people. It covers the entire region to collect a huge set of requirements. It is generally used for collecting requirement for general purpose software.

\subsubsection{Collaborative Techniques}
\textbf{Focus Group} is a technique where a group of four to nine users from different backgrounds, with different skills discuss in a free form, and concerns about features of a system that will be created.

\textbf{Brainstorming} provides an open environment of discussion, where users are free to give their requirement and expectation of the system. The data (ideas) collected after this process is then discussed and analyzed.

\textbf{JAD} (joint application development) is a requirement definition and software system design methodology in which stakeholders, subject matter experts (SME), end-users, software architects and developers attend intense off-site meetings to work out a system's detail. JAD focuses on the business problem rather than technical details.

\textbf{Prototyping}: A prototype of a system is an initial buildup of the system which often used to validate system requirement. There are two different types of prototypes: Throw-away prototypes help to understand difficult requirement. Evolutionary prototypes deliver a working system to the customer and often become a part of the final system.

\textbf{Work shop} is a collection of different types of stakeholders together to collect requirements for the project being developed. The workshop provides the complete set of requirement. It is very useful to elicit requirement for complex and large system.

\textbf{Story boarding} uses images, text, audio, video, animation diagram to visualize the concept to the stakeholders. This technique is allowing the stakeholders to come into common understanding of about the functionality of the system being developed.

\textbf{Models} include diagram such as Data flow diagram, Statechart, UML diagram to elicit requirement. The models use for the purpose to help the customer to thought the process. Models are useful for eliciting requirement and to resolve conflict between stakeholders.

\textbf{Use cases/Scenarios}: Use cases describe interactions between the user and the system to find user need. It specifies a sequence of interaction between a system and an external actor. Use cases represent functional requirements of the software system. \textbf{Scenarios} should include a description of the state of the system before entering and after completion of the scenario, what activities might be simultaneous, the normal flow of events and exceptions to the events.

\subsubsection{Cognitive Techniques}
\textbf{Document Analysis} is the process of analyzing the documents related to the problem domain to gather the information which is flowing in the organization. It is a useful technique to find in-depth knowledge about a particular task.

\textbf{Card sorting}: The card sorting method is used to generate information about the associations and grouping of specific data items. Participants in a card sort are asked to organize individual, unsorted items into groups. Card sorting may be conducted as a series of individual exercises, as a concurrent activity of a small group, or as a hybrid approach where individual activity is followed by group discussion of individual differences.

\textbf{Protocol analysis} is a method of conducting a meeting where stakeholders and analyst discuss the requirements of the system. Protocol analysis also provides the required actions to be taken for fulfilling the user requirements by using rationale.

\textbf{Laddering} is a form of structured interview in which a limited set of standard questions is asked to stakeholders. The set of questions is arranged in hierarchical order. The requirement for the success of this technique is that stakeholders have the domain knowledge.

\textbf{Repository grid} is a technique of developing a grid of the form of a matrix store the requirements involve asking stakeholders to develop attributes and assign values to a set of domain entities. It is a good technique to provide the distinction between different information domains.

\subsubsection{Observational Techniques}
\textbf{Observation}: Observational methods involve an investigation of user's work and taking notes on the activities that takes place. Observation may be either direct or indirect. Observation allows the observer to view what users actually do in context, overcoming issues with stakeholders, describing idealized or oversimplified work processes.

\textbf{Ethnography/Social analysis} is the process of interacting with stakeholder and user of different background to find out political environment within the organization. The observers go through the in depth observation of the organization to understand working and cultural environment.

\section{Data Collection - Matrices}
In this section, we present the strategy that we applied to select papers for populating the matrices. We initiated our search by identifying a query string being used to perform electronic searches, based on the 3PM. Then, we searched four electronic databases using this query string (IEEE Xplore, Springer, ACM digital Library, Science Direct). In addition, as a complement to the electronic search, we performed manual search in specific journals and conference proceedings, and also manually checked Software Engineering textbooks. We then scanned all the sources resulting from this two-stage search to select the works to be included in the review. During this step we applied
inclusion/exclusion criteria to select the techniques. For each paper, we read the paper's title and abstract to see whether it was relevant to our research topic. If the title and abstract of the paper could not help us make a decision, we further checked the paper's full text. In order to augment our collection of primary studies, we scanned the reference lists of all the identified primary studies to identify additional papers. Furthermore, we also went through publication lists of primary studies' authors to make sure that the most recent  publications on the same or similar topics were included.

In order to select an elicitation technique(s) for yielding an optimum result in the requirement elicitation process, we examine the merits of the elicitation techniques with respect to the project, process and the people of the software under development. Here, we collected the information about the stated parameters from the research papers [10-11][19][39][41-42][45][47-48][50][52][54-55][61-63][66-70][72], experience reports [36-37][40][57-59][71], white papers [38][51][58][56][65] and the project documents [44][35][46] to provide the methodological guidance about the selection of the techniques [43][49][53]. This helps us to create the matrices for all the three parameters. The research approach utilizes the extracted information about the software project to identify the set of elicitation techniques. This helps the analyst to gather quality requirements from the stakeholders that would result good quality software. Getting a subset of techniques from the larger available set of elicitation techniques will help the system analysts to concentrate only on the identified techniques, instead of, choosing the technique on the basis of personal experience of analysts. Choosing the right technique, for the right project, and at right time, leads to quality software. Since, last 60 years of experience in software development, about the success or failure of the software project. It seems that, if we collect the information about the project success, failure, methods, techniques applied, dos and don'ts etc., we would achieve a very strong knowledge database of the software. This database would be very helpful in achieving the goals in context with the software and might be, we are in the positions to minimize the error/bugs in the software development life cycle.

\section{Framework Overview}
In this section, we present our proposed framework for the selection of requirement elicitation technique. This framework provides us the guidance to choose the requirement elicitation techniques in context with the affected parameters of the software i.e., people, process and the project attributes. As we know, requirement elicitation is performed in a wide variety of conditions, which include many dimensions representing various combinations of participants, problem domain, and organizational context.

\noindent \textbf{Assumptions}\\
First, let us consider the basic assumptions for successful execution of our approach.
\begin{itemize}
  \item A team of requirement analysts is available for performing the task, of applying the proposed approach.
  \item Identifying various important attribute in three dimensions namely project, people and process of the software development by the analysts for the software project being undertaken.
\end{itemize}

\noindent The framework are in four-folds:
\begin{enumerate}
  \item Identify various important attribute in each of three dimensions namely project, people and process of the software development separately.
  \item We construct three p matrix (3PM) separately for each dimension of the software and their relationship with the elicitation techniques and use them in the selection of one or more elicitation technique for a given software project to be undertaken.
  \item Extract the information about the elicitation techniques on the basis of 3P parameters.
  \item Provide a mapping mechanism to choose the set of elicitation techniques, to gather the requirement on the basis of 3PM and knowledge base for analysts to make a decision regarding selection of elicitation techniques.
\end{enumerate}

\begin{figure}
\centering
\epsfig{file=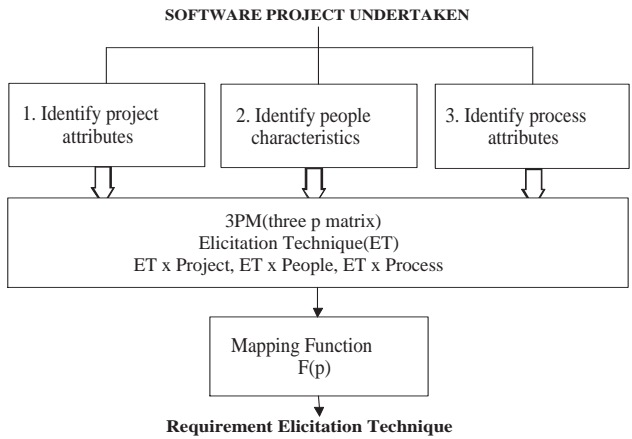, height=2.3in, width=3.3in}
\caption{Overview of the proposed framework}
\vspace*{-3.0ex}
\label{fig:fig1}
\end{figure}

In order to achieve these objectives, we have developed 3P matrix and a methodology within the framework for the selection of requirement elicitation technique. The proposed framework is given in Figure~\ref{fig:fig1}.

\subsection{1P: Project Matrix}
\begin{figure}
\centering
\epsfig{file=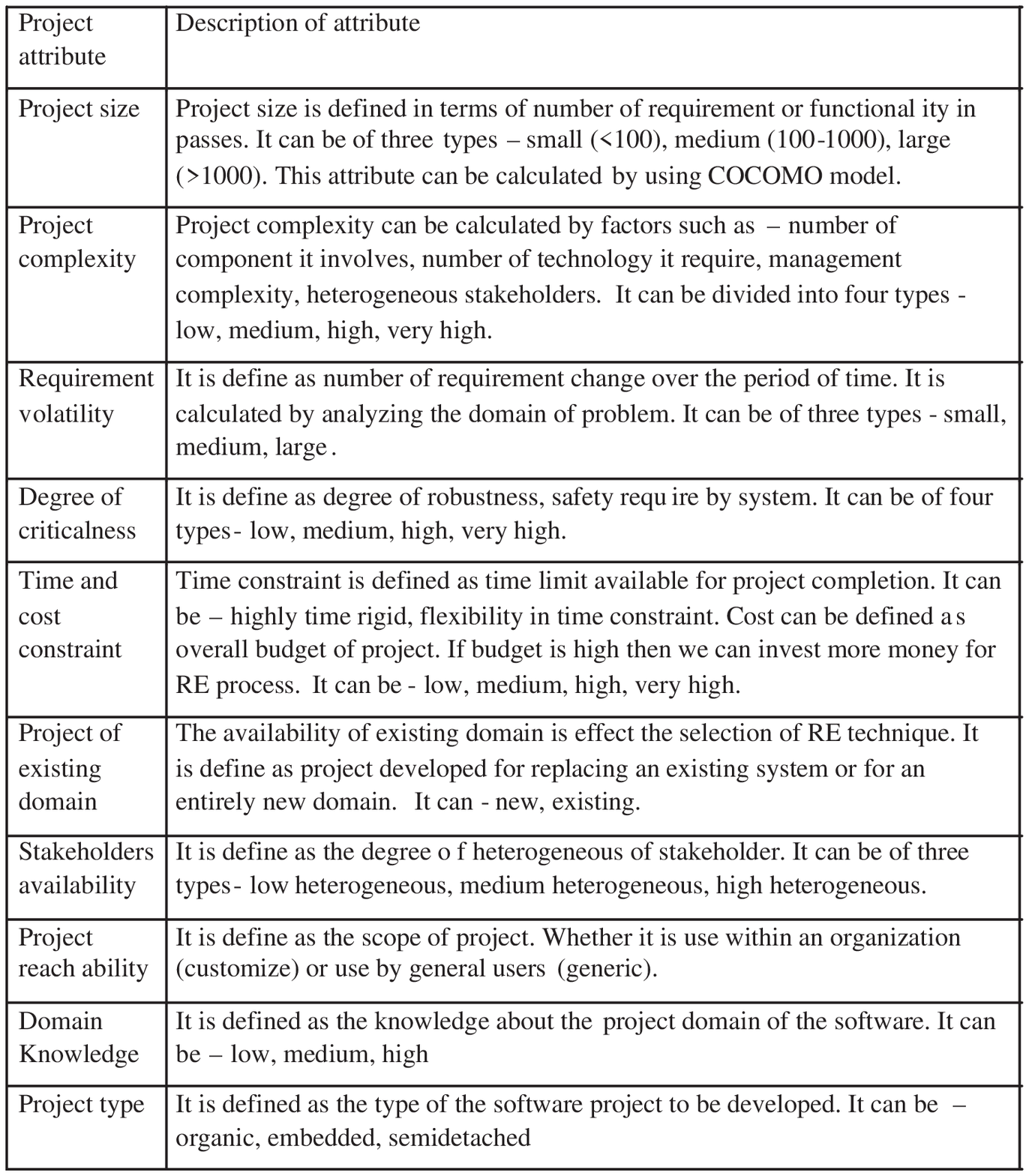, height=4.2in, width=3.5in}
\vspace*{-6.0ex}
\caption{Project attributes with their descriptions}
\vspace*{-4.0ex}
\label{fig:fig5}
\end{figure}

In this subsection, we define a set of software project attributes for a given software product. The project attributes play an important role when it comes to the project characterization and selection of elicitation techniques. Figure~\ref{fig:fig5} shows the set of project attributes used in our study for the selection of RE techniques. The purpose of this step is to understand the basic characteristics of the project and score the attributes as per the required elicitation technique. For eliciting the requirements of the user, the important issue is to incorporate the project attributes of the software project going to be developed, to choose the elicitation techniques. Here, we have developed a relation matrix between the project attributes and available elicitation techniques. This helps us in taking a decision to choose the appropriate elicitation techniques for the software project being developed. Each of these project attributes are shown in Table~\ref{fig:fig2} in relation with the set of elicitation techniques.

\begin{table*}[ht]
\caption{Project attributes matrix}
\centering
\begin{tabular}{cc}
\epsfig{file=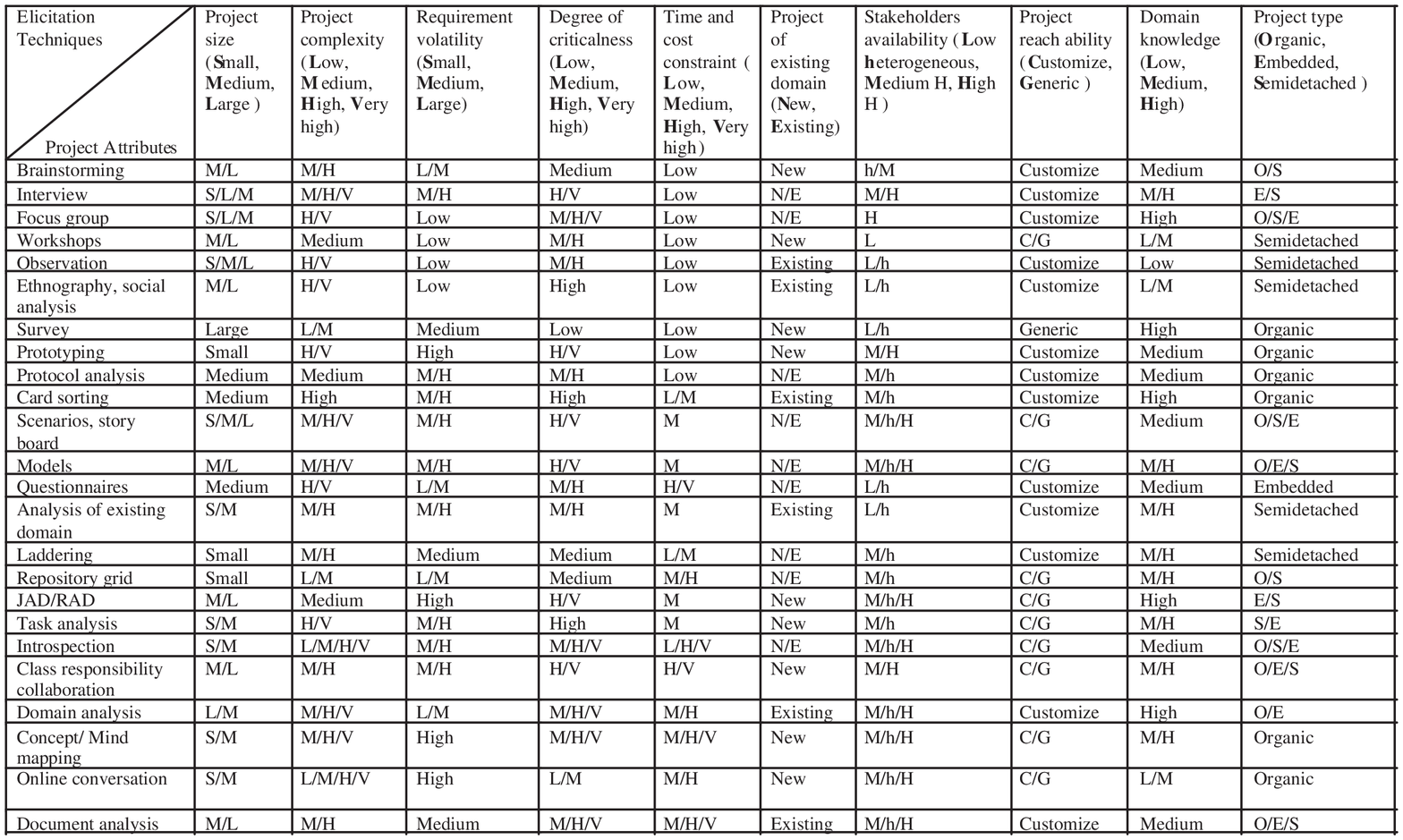, height=4.2in, width=7.2in}
\end{tabular}
\label{fig:fig2}
\vspace*{-9.0ex}
\end{table*}

\subsection{2P: People Matrix}
Moreover, the people involvement plays an important role in the software development process and also in the selection of elicitation techniques. The people behavior, knowledge, skill, experience and other parameters might affect the software development activity. The purpose of this step is to understand the basic characteristics of the people in accordance with the software project being undertaken. This matrix helps us to choose the elicitation technique on the basis of people attributes. In the people matrix, the 'N' represent No and 'Y' represent Yes. The relation between the elicitation technique and the people attributes are shown in Table~\ref{fig:fig3}.
\begin{table*}[ht]
\caption{People matrix}
\centering
\begin{tabular}{cc}
\epsfig{file=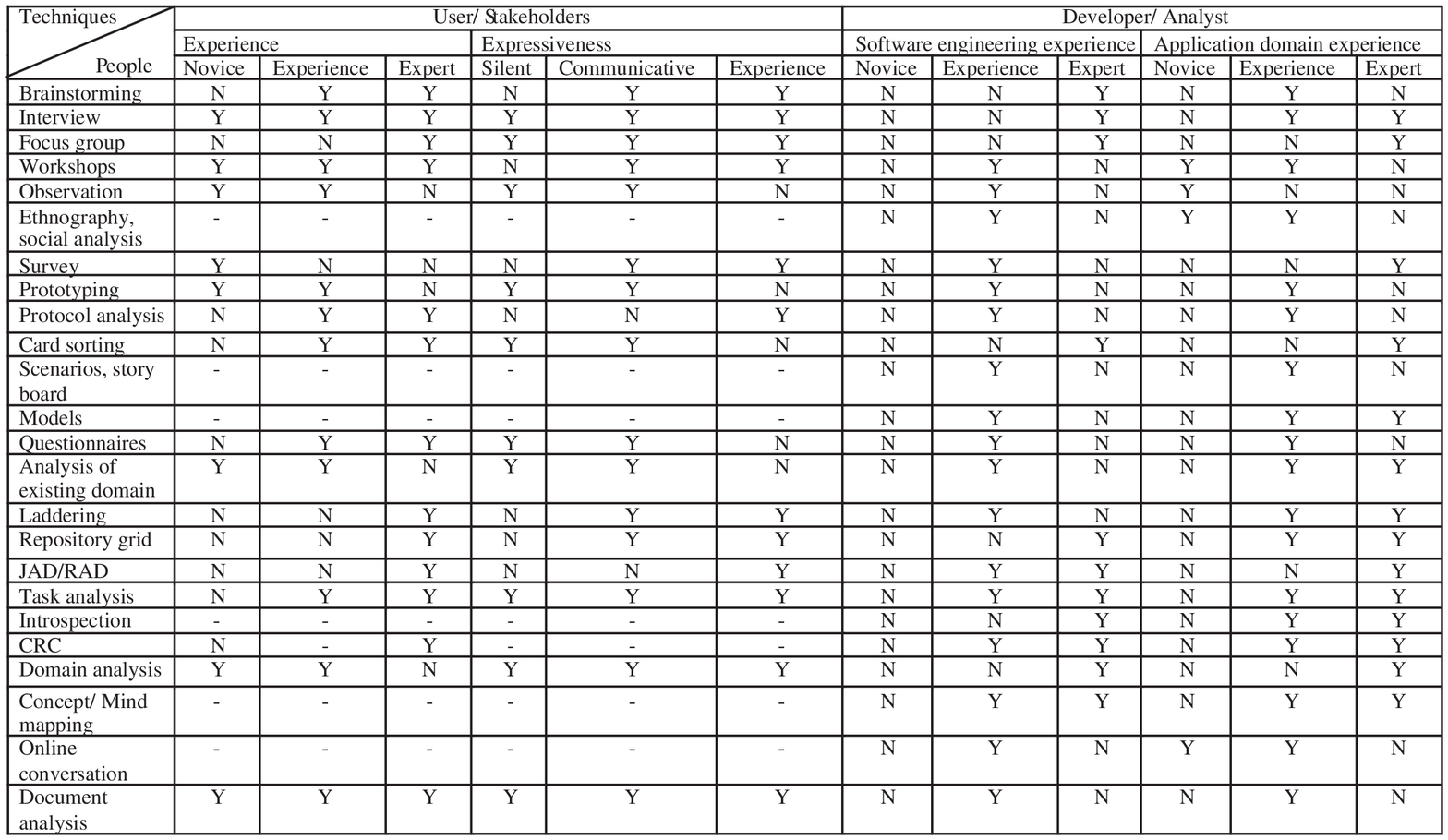, height=4.5in, width=7.1in}
\end{tabular}
\label{fig:fig3}
\vspace*{-9.0ex}
\end{table*}

\subsection{3P: Process Matrix}
A set of software processes was defined in the books and research papers for effective and efficient software development. The processes involved in the software development may play an important role, when it comes to the selection of elicitation techniques. Each of these attributes of software processes are shown in Table~\ref{fig:fig4} in relation with the set of elicitation techniques. In this process matrix, we have filled the values between 0 to 1, on the basis of required score, the corresponding technique is used for that software process. The values specified in the process matrix are based on the available knowledge from the reports, white papers, research papers etc. The reason to specify these values in number, due to the nature of software processes as they can be applied in any software irrespective of its domain. There exist two cases when analyzing the results of the process matrix i.e., \textbf{Case 1:} If the score of elicitation technique is less than 0.5 (Score$<$0.5), then it is decided not to use them for that process model and \textbf{Case 2:} If the score of elicitation technique is greater than or equal to 0.5 (Score$\geq$0.5) then, it is an effective technique to use for that process model.
\begin{table*}[ht]
\caption{Software process matrix}
\centering
\begin{tabular}{cc}
\epsfig{file=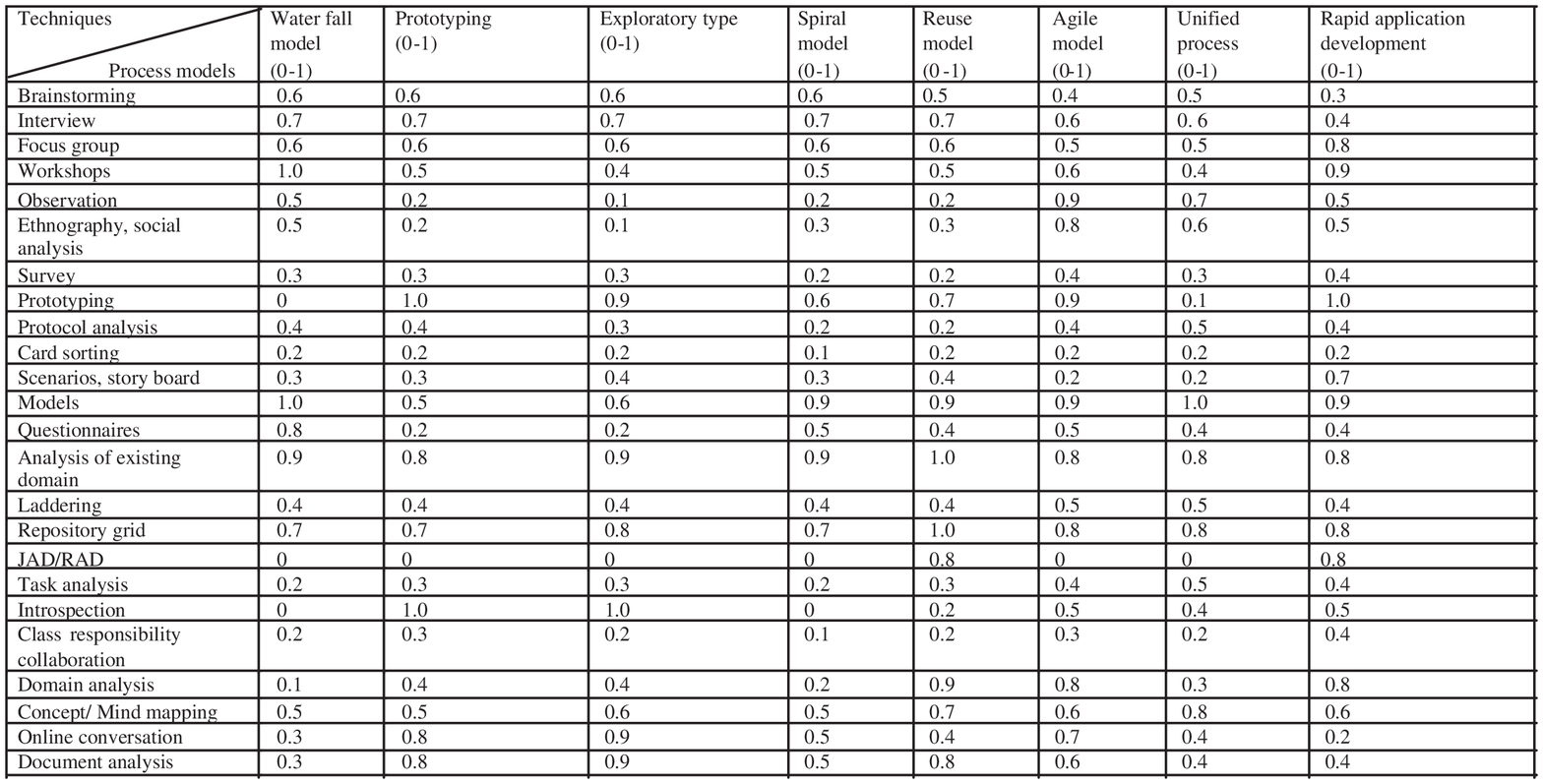, height=3.8in, width=7.0in}
\end{tabular}
\label{fig:fig4}
\vspace*{-9.0ex}
\end{table*}

\subsection{Mapping Criteria}
In order to apply the combination of all three p matrices for a given software project and, to identify the set of elicitation techniques, a systematic methodology is needed to ensure success. We, therefore develop a mapping function for requirement elicitation techniques selection. Let us consider the three p matrix (3PM) $A_{ij}$, $B_{ik}$ and $C_{il}$ respectively where, i represents the set of elicitation technique, j represents the set of project attributes, k represents the set of people characteristics and l represents the set of software development process models.

The criteria used for selecting the subset of elicitation technique, on the basis of three P matrices are as follows.
\begin{enumerate}
  \item We divide the undertaken software project into the attributes of project, people and process. This is done according to the characteristics of the software, as identified by the requirement analysts during the feasibility study phase of software development (This is a very first phase before requirement elicitation process).
  \item Next, we identify the subset of techniques from each p matrix by simply analyzing them i.e., For $A_{ij}$, for each project attribute we get some set of elicitation techniques. To get the subset of elicitation technique for the $A_{ij}$ matrix, we do the union of techniques  identified corresponding to each attribute.
  \item Next, we do union of all the techniques determined in each p matrix. Step 1 and 2, resulted a subset of non-repeated elicitation techniques i.e.,
      Set of Elicitation technique in $A_{ij}$ $\cup$ Set of Elicitation technique in $B_{ik}$ $\cup$ Set of Elicitation technique in $C_{il}$ $\longrightarrow$ Final subset of elicitation technique using 3PM according to the software project being undertaken.
  \item Next, we perform feasibility study from user as well as analyst (elector) points of view to determine the set of techniques that are actually applicable during the elicitation process.
\end{enumerate}

\section{Case Studies}
This section shows the applicability of the proposed approach using several case studies. We apply the approach to an IPOS [55] (Intelligent Power Optimization System) project, Online shopping mall project and Bhoomi E-governance project. First project is an intelligent power optimization system, second is a web based portal of a shopping mall and last one is a government project which provide information of about roads and lands of a state. Moreover, presented case studies have been conducted throughout the course of this research. In these case studies, we analyze the project attributes, software processes and people characteristics of each project. Next, these identified characteristics are used as an input in our framework to develop a relationship between the requirement elicitation techniques and 3P parameters corresponding to the software projects. For these case studies we make some assumptions based on our best knowledge to find the parameters of the project.

\subsection{IPOS Project}
\subsubsection{Project Description}
This project is a large sized project to develop an Intelligent Power Optimization System (IPOS). A set of elicitation techniques are selected by requirement analysts by applying the proposed approach. The selected techniques were considered, the most suitable techniques for the project IPOS. The following subsections give a brief description of the case study and the major findings.

\subsubsection{Technique Selection and their Usage in Project}
First, we identify the project attributes of the IPOS project i.e., (i) Project Size: Large (between 1000 to 4000) (ii) Project Complexity: Very High (iii) Requirement Volatility: Very Low (iv) Degree of Criticalness: High (v) Time and cost constraint: Low (vi) Project of existing domain: Existing and so on. Using the project attribute matrix, we identified that Focus group, Interview, Ethnography are the recommended techniques.

Next, we identify the people attributes of the IPOS project i.e., User/Stakeholders: Novice, experience, communicative; Developer/Analyst: Novice, experience. Using the people attribute matrix, we identified that Observation, Interview, Focus group are the recommended techniques.

Similarly, for the process attributes of the IPOS project, i.e., for the agile process model, we get the following techniques: Interview, focus group, workshops, observation, ethnography, prototyping, models etc.

\subsubsection{Results}
Finally, using the mapping criteria we suggest the small subset of elicitation techniques for the IPOS project. The recommended techniques are Interview, focus group, workshops, observation, ethnography, prototyping, models.

\subsection{Online shopping mall project}
\subsubsection{Project Description}
The Online Shopping Mall (OSM) web application is intended to provide complete solutions for vendors as well as customers through a single get way using the internet as the sole medium. It will enable vendors to setup online shops, customer to browse through the shop and purchase them online without having to visit the shop physically. The administration module will enable a system administrator to approve and reject requests for new shops and maintain various lists of shop category.

\subsubsection{Technique Selection and their Usage in Project}
First, we identify the project attributes of the OSM project i.e., (i) Project Size: Medium (between 100 to 1000) (ii) Project Complexity: Medium (iii) Requirement Volatility: Medium (iv) Degree of Criticalness: Medium (v) Time and cost constraint: Low (vi) Project of existing domain: New and so on. Using the project attribute matrix, we identified that Brainstorming, Focus group, Interview, Ethnography, observations, models, questionnaire are the recommended techniques.

Next, we identify the people attributes of the OSM project i.e., User/Stakeholders: Novice, Silent; Developer/Analyst: Novice, experience. Using the people attribute matrix, we identified that Observation, Interview, Focus group, prototyping are the recommended techniques.

Similarly, for the process attributes of the OSM project, i.e., for the prototyping process model, we get the following techniques: Interview, focus group, workshops, prototyping etc.

\subsubsection{Results}
Finally, using the mapping criteria we suggest the small subset of elicitation techniques for the OSM project. The recommended techniques are Interview, focus group, workshops, observation, ethnography, prototyping.

\subsection{Bhoomi E-governance project}
\subsubsection{Project Description}
The purpose of Project Bhoomi is to present a highly scalable, extendable, robust, user-friendly, easily deployed and cross-platform prototype for a software system for managing the City Land Record Management System of a State. The objective is to provide free flow of information and better
governance through the use of technology (e-Governance). A person should be able to access the desired information anytime-anywhere. At the administration/governance side, the software solution should provide easier and effective editing (addition/removal/change) of records and a better control and monitoring of the land records, apart from the requirements above.

\subsubsection{Technique Selection and their Usage in Project}
First, we identify the project attributes of the Bhoomi project i.e., (i) Project Size: Large (between 1000 to 4000) (ii) Project Complexity: Very High (iii) Requirement Volatility: Very Low (iv) Degree of Criticalness: High (v) Time and cost constraint: Low (vi) Project of existing domain: Existing and so on. Using the project attribute matrix, we identified that Focus group, Interview, Ethnography, Observation, Models, Survey, Introspection are the recommended techniques.

Next, we identify the people attributes of the Bhoomi project i.e., User/Stakeholders: Novice, experience, expert, communicative; Developer/Analyst: expert, experience. Using the people attribute matrix, we identified that Brainstorming, Observation, Interview, Focus group etc are the recommended techniques.

Similarly, for the process attributes of the Bhoomi project, i.e., for the waterfall process model, we get the following techniques: Brainstorming, Interview, focus group, workshops, observation, ethnography, models, questionnaire, analysis of existing domain, concept/mind mapping etc.

\subsubsection{Results}
Finally, using the mapping criteria we suggest the small subset of elicitation techniques for the Bhoomi project. The recommended techniques are Brainstorming, Interview, focus group, workshops, observation, ethnography, models, questionnaire, analysis of existing domain, concept/mind mapping, survey.

\section{Comparison with Other Approaches}
This section presents the validation of the proposed approach in comparison with other approach presented by [55][56]. Table~\ref{tab:tab2} shows the comparison between the techniques where `Y` represents the presence of property. In the work reported, [55] presented their experience of improving the requirements engineering process for a software project using a combination of Requirement Engineering (RE) techniques based on project attributes and characteristics of RE techniques. The case study of IPOS is conducted which shows a positive result of the application of a combination of RE techniques to a software project using only project characteristics. Similarly, we have also included a case study of IPOS to apply the proposed approach using all the three matrices. The results identified (see Section Case Studies) suggests that the populated matrices are useful for selecting the subset of requirement elicitation techniques on the basis of project, people and process attributes.

Jiang et al. [56], presented a methodology for RE process development for a given project where first, a RE Process Knowledge Base (REPKB) is established.  Second, a decision support mechanism is provided during RE process development. Third, this methodology uses three components: process building blocks, standard templates of the RE process and development guidelines, to help process development. Fourth, it explicitly links project characteristics with RE process development so that the most suitable RE process can be developed. Similar, to this approach we have also generated the three matrices to acquired the knowledge about the elicitation techniques. In their approach, they had only considered the project attributes, while others are not considered.

\begin{table}[!ht]
  \centering
  \caption{Comparison}
    \begin{tabular}{|l|c|c|c|} \hline
    \#    & [55]  & [56]  & Our Approach\\ \hline
    Project Characteristics & Y & Y & Y \\ \hline
    People Attributes & - & - & Y  \\ \hline
    Process Attributes & - & - & Y  \\ \hline
    Effective Approach & Y & Y & Y  \\ \hline
   \end{tabular}
  \label{tab:tab2}
  \vspace*{-2.5ex}
\end{table}

\section{Discussions}
Identify the set of elicitation techniques for the software project to be undertaken is an important task to gather quality requirements. In this paper, we have proposed an approach where we identify some initial characteristics of the software before the process of requirement gathering begins. These initial characteristics are project attributes, people attributes and the process model involved in the software. The reasons to include all these parameters are:
\begin{itemize}
  \item The requirement gathering process is a human centric process. Thus, it needs to include the behavior or skills of the people that directly or indirectly involved with the software project.
  \item The software project is influenced by several parameters. Therefore, it is important to include the project attributes in our study.
  \item The process models involved in the development of software, needs some set of elicitation techniques according to the behavior. So, we have also included the process models as an important parameter for the study.
\end{itemize}

This information will be extracted with the help of project managers, requirement analysts to identify these above stated parameters for the software project. The obtained information and available information in the matrices database are combined with the available techniques to identify the set of elicitation techniques. Moreover, it is not mandatory to use all three matrices together for getting optimum elicitation technique. They can be used separately depending upon the type of software project being undertaken, or on the basis of industry experience. The information/values stated in the matrices might be used by researchers or industry persons to populate, for further detailed study.

The limitations of our research are influence by the scope of elicitation technique which we included and our understanding of these techniques. The state-of-the-practice in RE is still one of the major problems in software development; it is very difficult to reach to any conclusion for selecting requirement eliciting techniques. But by analyzing the domain of problem we can find some characteristics of the system, which can help us to making decision of selecting requirement eliciting techniques.

The proposed approach has some limitations. The approach presented is a heuristic because we have taken some assumptions to identify the requirement gathering techniques due to the lack of information or guidance related to requirement engineering. We have taken the help of various research papers to maintain the database and for filling the relevant information required to gather the requirement. Although, we know that the mapping function used by the approach is theoretical one, but it might be helpful to the analysts for choosing the set of best elicitation techniques.

\section{Related Work}
The selection of suitable RE techniques for a specific domain of software project is a challenging issue. Number of variables influences this selection process. To overcome this challenge, several solutions have been proposed by many authors in different perspectives: Method engineering~\cite{rf:kumar}\cite{rf:Brinkkemper}[5][6][7][8][20] provides approaches to develop or adapt existing methodologies to the new problem domain. Maiden et al. [9] proposed a framework which provides guidance to selecting technique for requirement elicitation. Hickey et al. [10][11] proposed a model that helps to understand the RE process and the selection of elicitation techniques. Bickerton et al. [12] provide a framework for the classification of elicitation techniques based on the social assumptions. Macaulay [2] proposes a list for the selection of elicitation techniques. Kotonya et al. [13] proposed characteristic attributes which can be helpful for the selection of elicitation techniques. Davis et al. [14] also proposed a process for elicitation technique selection based on four strategies of the requirements determination model and focuses on the selection of RE elicitation techniques. Browne et al. [15] also proposed guidance for the selection of elicitation technique. Lobo et al. [16] developed an approach for elicitation technique selection based on a predefined RE process model. Lausen [18] discusses several techniques, which can be used in the RE process. He also provides an idea for elicitation technique selection. After of many study and research, the problem of RE is not address properly and there is still lack of proper contextual information for selecting elicitation is available.

Viviane et al., [26] proposed an collaborative approach for requirement elicitation process. There approach consists of a knowledge model based on the stories about the system and a tool to support interaction. Suranjan et al., [27] describe an approach to provide a guidance for the process of requirement engineering. Their study collaborates based on the four states and identifies important factors that tends to trigger from one state to another state. Ruben et al., [28] provides a framework based on the need of modeling primitives and it supported by the mean of theoretical and modeling foundation of a social science framework. Yan Tang et al., [29] proposed a framework to provide the guidance for the selection of requirement elicitation technique. Bee Bee et al., [30] proposed guidance for understanding the use of elicitation approach for effective requirement gathering. Sumaria et al., [31] introduces a way and the guidance for selecting requirement elicitation techniques. They analyzed different elicitation techniques in the context of different project settings. Zhying [32] provides a comparison of different requirement elicitation techniques and concludes his findings.

Ganesh et al., [60] presented literature study and the experimental case study on analyzing and compare different methods for requirement gathering process, this provides the flexibility to requirement engineers to know the characteristics and effectiveness of every method. All these information might be useful to select the particular elicitation method depends on the type of application and the situation. On the other hand, due to lack of guidance on which techniques are suitable for a certain project context it is difficult to choose the set of elicitation technique. Thus, Jiang et al., [64] proposed a methodology for Requirements Engineering Techniques Selection (MRETS) as an approach that helps requirements engineers select suitable RE techniques for the project at hand.

In this paper, we focus our effort to identify the characteristics of any problem domain in the context of project attributes, people and process of software development and then try to find the relationship between characteristics and available techniques to provide a guidance of selecting requirement elicitation technique. Then, we would be able to improve the average analysts abilities to select proper elicitation technique.

\section{Conclusions and Future Work}
This paper has attempted to present meaningful insights into the feature of different types of requirements elicitation techniques. As discussed above that RE process is highly influence by three parameters namely people characteristics, involved processes and the project attributes. Each elicitation technique has a predefine set of specific and unique characteristics and their context of application. In this paper, we present an approach to select a subset of requirement elicitation technique for an optimum result in requirement elicitation process. The proposed approach first, identifies the attributes in three dimensions namely project, people and process of the software development, that can influence the outcome of an elicitation technique. Second, we construct three p matrix (3PM) separately which shows a relation between elicitation techniques and three dimensions of a software. Finally, we provide a mapping criteria and use them in the selection of elicitation techniques.

We aim to extend this work by incorporating the industries data in 3P matrices and subsequently, analyze these matrices by applying the organization specific case study for the selection of elicitation techniques.

\bibliographystyle{IEEEtran}
%

\end{document}